# Radioactivity in a bucket

Luis Peralta[1,2]

[1] Faculdade de Ciências da Universidade de Lisboa, Lisboa, Portugal
[2] Laboratório de Instrumentação e Física Experimental de Partículas, Lisboa, Portugal

E-mail: luis@lip.pt

**Abstract**

In Radiation Physics classes, point sources are typically used, for which it is relatively easy to describe the signal obtained by a radiation detector, such as the NaI(Tl) scintillation detector. The use of large extended radiation sources is generally avoided due to the mathematical complexity that their description may involve. However, the use of Monte Carlo simulation methods allows this limitation to be overcome. Potassium chloride, containing the $^{40}$K isotope, is an ideal candidate for carrying out this type of experiment. The source activity is obtained through the detection of the 1460.8 keV gamma- photon emitted in the $^{40}$K decay. In the first experiment, a cylindrical container is used, placing the NaI(Tl) detector in its center and filling the remaining space with potassium chloride. In a second, more complex case, a large radioactive source consisting of a container filled with a mixture of sand and potassium chloride, with the NaI(Tl) detector placed in the center of the mixture, is used. In this case, the mass of potassium chloride is approximately 1/5 of the sand mass. In both experiments, the detection efficiency is obtained by Monte Carlo simulation. A careful analysis of the experimental data allows to obtain a good agreement between the measured and calculated value of the activity.

## 1. Introduction

There are many examples of large radioactive sources that we can find. As examples we have the atmosphere where we find radon gas and its radioactive progeny or the concrete floor of a building that contains, among other elements, traces of potassium. As the determination of the activity of a source scattered in space is considerably more complex than the determination of a quasi-point source, the subject is generally not addressed in books dedicated to Nuclear or Radiation Physics. In general, the activity determination of a large radioactive source will depend on multiple factors such as geometry, detector to source distance, radiation absorption in materials between the source and detector, etc. Although complex, the subject can be approached experimentally and computationally by university students.

In this work we built a model of a large radioactive source where a quantity of potassium chloride was mixed with river sand. The experiment goal is to determine the source's activity and from that measurement to determine the mass of potassium chloride dispersed in the sand. As a comparison term, the activity determination of a KCl source of known mass was also made.

## 2. The experimental setup

In the experiments, the gamma radiation of 1460 keV emitted by the $^{40}$K isotope is detected by a 7.62 cm × 7.62 cm (3"×3") NaI(Tl) Canberra detector. The preamplifier output of the detector was connected to an Ortec 575A amplifier and the signal was digitized by an Amptek MCA8000A multichannel analyzer. The source material was placed in plastic buckets, with an aluminum tube placed in the center, which will contain the NaI(Tl) detector. The aluminum tube used is of the type used in gas exhaust installations, such as domestic water heaters and is easily found in hardware stores. The source material was subsequently placed around these tubes. In this way it is possible to easily remove the detector after the measurement is made. Three buckets containing different materials were prepared. A bucket with just potassium chloride, a bucket with a mixture of sand and potassium chloride, and a bucket with just sand. Due to the considerable weight of the latter buckets, these were placed on a base with wheels, so that they could be moved from the radioactive source vault where they are kept to the laboratory. Figure 1 shows photographs of the three containers. Container 1 (the smallest of the three buckets) just has KCl. The KCl sample forms a ring with a hole in the middle for insertion of the detector. In the classroom experience all the characteristics of this sample are provided to the students. It is intended in this way that they are able to validate the entire detection process using an example in which they know the expected result. The mixture of KCl and sand intends to simulate the situation of a deposit of radioactive material. An approximate ratio of 1:5 was used. Since sand also contains potassium, it is necessary to measure the activity per unit mass of a sample of sand. This will be the function of the bucket containing only sand. The size of container 3 being different from container 2, is a deliberate additional complication for students to sort out the correct answer. The dimensions and total masses of both buckets are given to students. However, the mass of KCl that was mixed in the sand is not revealed, being asked to determine it from the measurements made.

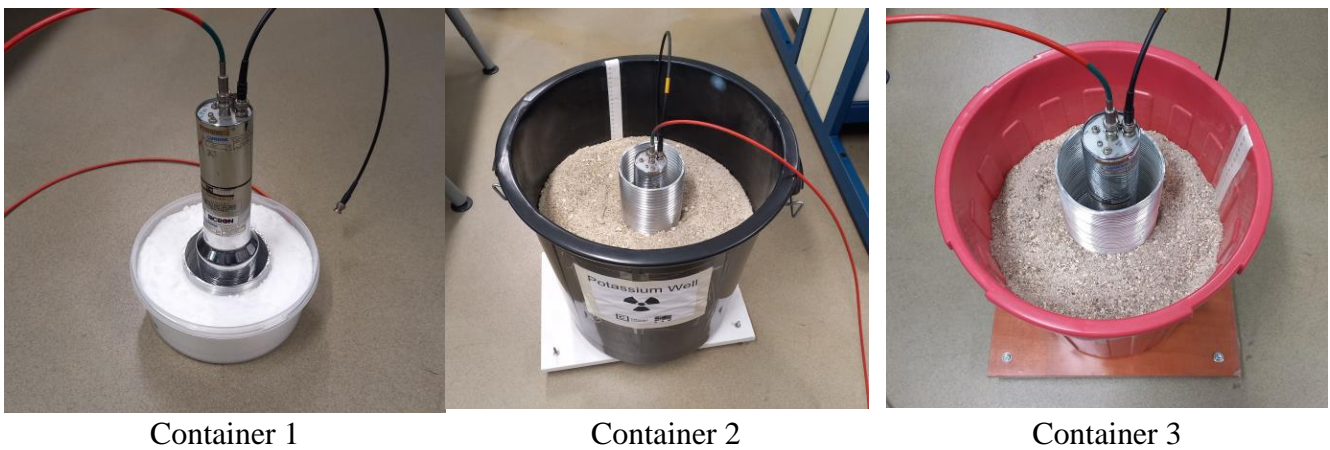

Container 1     Container 2     Container 3

Figure 1. Photo of the containers with the radioactive sources under study. From left to right: container 1- KCl; container 2- sand+KCl; container 3- sand.

Table 1 shows the mass values of the components and dimensions of the containers used. The density values presented result from direct measurements carried out in the laboratory on samples of each of the materials. In particular the KCl sample density differs from the standard value of 1.984 g/cm$^3$ [1] due to its powder form. The sand+KCl and sand sample containers are slightly conical and average diameter values are given in the table.

Table 1: Mass of samples and dimensions of containers.

| Container | Source | Mass (kg) | Density (g/cm$^3$) | Average diameter (cm) | Source height (cm) |
|---|---|---|---|---|---|
| 1 | KCl | 2.598 ± 0.002 | 0.93 ± 0.02 | 24.0 ± 0.2 | 7.6 ± 0.2 |
| 2 | Sand KCl | 37.235 ± 0.010 7.447 ± 0.005 | 1.25 ± 0.02 | 40.5 ± 0.2 | 30.0 ± 0.5 |
| 3 | Sand | 21.613 ± 0.010 | 1.56 ± 0.02 | 28.3 ± 0.2 | 23.0 ± 0.5 |

The masses of sand and KCl were measured on a balance with a resolution of 0.01 g. The quoted uncertainties take into account the fact that the masses were divided into several fractions before being measured, as well as possible systematic errors made during the measurement. The uncertainty in the average diameter is an estimate made based on measuring the largest and smallest diameter of each container with a ruler. The uncertainty in the height of the masses in each container takes into account the fact that the free surface (ie the upper surface) is not strictly flat.

A sand sample was analyzed by X-ray fluorescence. The percentage by weight of the main elements are shown in table 2.

Table 2: Percentage by weight of the main elements present in the sand sample.

| O | Si | Al | K | Fe | Ca | Mg | Ti |
|---|---|---|---|---|---|---|---|
| 43.5% | 38.1% | 4.7% | 3.2% | 1.4% | 0.87% | 0.45% | 0.36% |

In addition to the radioactivity due to the potassium contained in the sand in container 2, it is necessary to consider the background due to the potassium present in the concrete of the laboratory floor, so it will also be necessary to carry out an acquisition of this background.

In general, the activity $A$ of a radioactive source that emits gamma radiation can be obtained from the following equation [2]

$$A = \frac{N_{signal}/t}{\varepsilon B_r} \quad (1)$$

where $N_{signal}/t$ is the signal count rate, $\varepsilon$ the total detection efficiency to the emitted photon in the transition with branching ratio $B_r$. While the signal count rate is obtained from experimental measurements, the total efficiency for a non-trivial geometry case can be obtained by Monte Carlo simulation. The total detection efficiency for a given type of particle is defined as the ratio between the number of these particles that are detected by the total number of that type of particle emitted by the radiation source. Thus, the detection efficiency depends not only on the characteristics of the detection system (detector and acquisition system included), but also on the characteristics of the radiation source and the medium that the particles pass through until they hit the detector.

**3. Monte Carlo simulation**

The Monte Carlo simulation of the various experimental setups was performed with the Penelope program [3,4] coupled with the tracking and histogramming program Ulysses [5]. Figures 2 and 3 show the simulated geometries for each case. The NaI(Tl) detector is simulated as a cylindrical crystal 7.62 cm in diameter and 7.62 cm in height. The crystal is surrounded by a 0.185 cm thick MgO reflector and is enclosed in a 0.05 cm thick aluminum case [6]. In this simulation the glass on the back of the crystal is not considered since the simulation does not create the scintillation photons produced in the crystal. The sample volumes are simulated as cylinders. For each container a different chemical composition was considered in the simulation. The measured densities were used, notably in the case of the KCl sample, where the measured value departs significantly from the reference value [1]. Although in this work the measured sand composition was used, a simple composition of the sand as $SiO_2$ can also be considered without loss of the experiment accuracy. The NaI(Tl) detector is placed inside an air well. The source volume coincides in each case with the sample volume, and the emission of 1460 keV photons is isotropic. In order to reduce computation time, the Ulysses program checks if the photon flight line intersects the detector volume. If not, that photon is not followed, but still accounted for the total number of generated events. To simulate the resolution of a typical 3"×3" NaI(Tl) detector, a Gaussian dispersion is added to the value of the energy deposited in the crystal, where a *FWHM* of 7% for the energy of 662 keV [7] was assumed. A simplified model of the resolution $R$ as a function of energy E was considered

$$R = \frac{FWHM}{E} = \frac{a}{\sqrt{E}} \qquad (2)$$

where $a$ is a constant. For the assumed value of 46.34 keV for *FWHM* at 662 keV a value $1.80\ keV^{-1/2}$ is obtained for constant $a$. The scoring of the deposited energy in the crystal is made in a histogram. The total efficiency (including geometric efficiency, attenuation in different material media and intrinsic efficiency of the NaI(Tl) crystal) is obtained as the number of events obtained by simulation in the photopeak, divided by the total number of generated events.

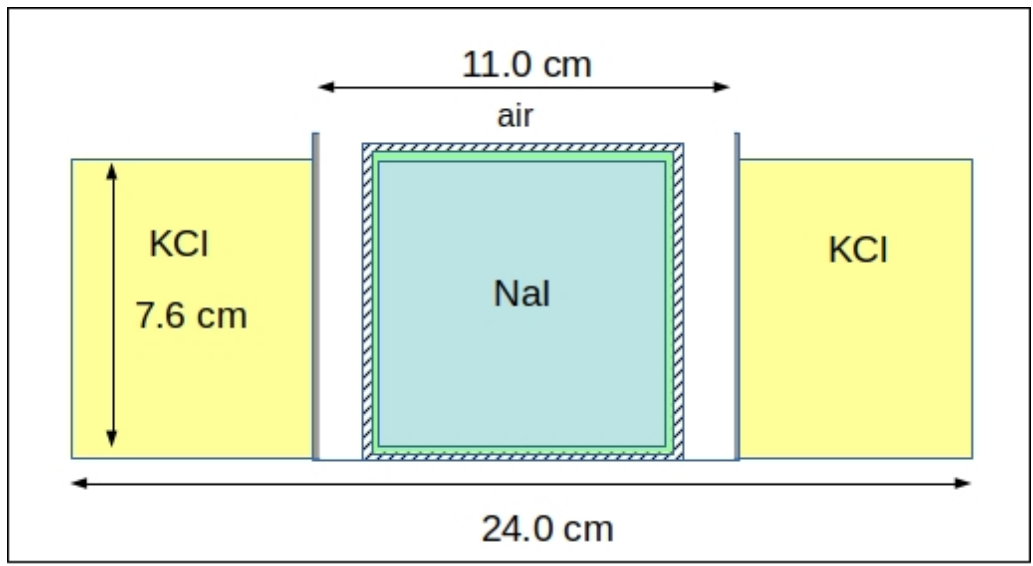

Figure 2. Cross-section of the geometry used to simulate the NaI(Tl) detector inside container 1. All volumes are simulated as 3D cylinders.

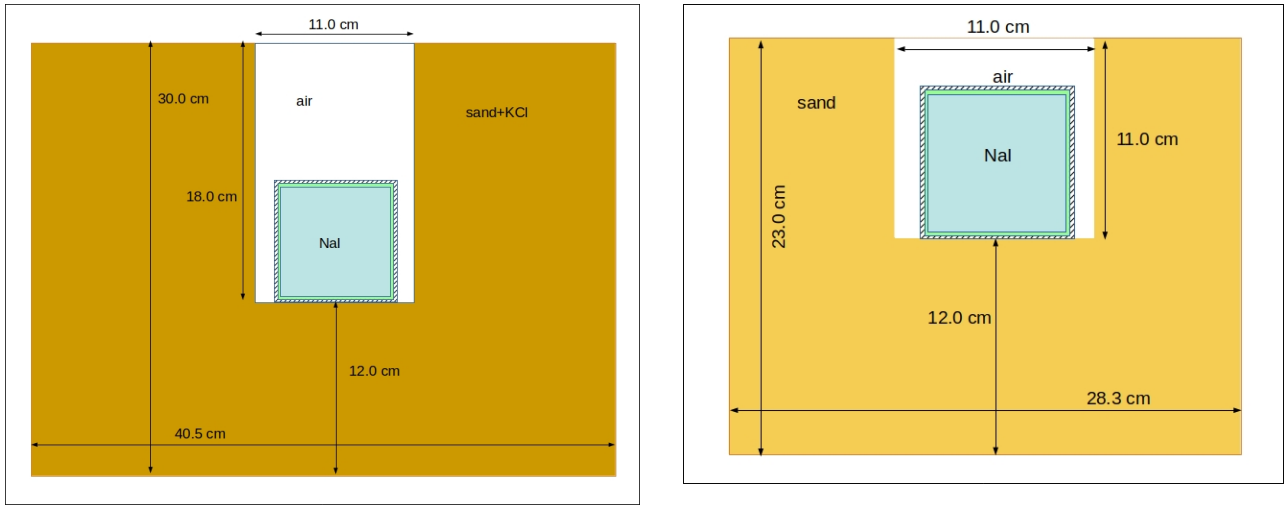

Figure 3. Cross-section of the geometry used to simulate the NaI(Tl) detector: left) located inside container 2; right) located inside container 3. All volumes are simulated as 3D cylinders.

The efficiency values obtained are presented in Table 3. The statistical uncertainties are relatively small, however there are systematic uncertainties that have to be estimated. The evaluation of the efficiency uncertainty due to uncertainties in the source dimensions and crystal size were made by simulating the setup with slightly changed dimensions varied according to the uncertainties presented in Table 1. Size variations of the crystal and container were done in a way to maximize the efficiency deviation relative to the previously obtained value. That was accomplished by decreasing the NaI crystal size and increasing the sample volume. These uncertainties are quoted as systematic and presented separately from statistical uncertainties in table 3. These uncertainties are also shown in Table 3, their value being considerably higher than the statistical uncertainty.

Table 3: Efficiency value for the detection of the 1460 keV photopeak, for each of the simulated situations. The first uncertainty value is statistical and the second is the systematic uncertainty value.

| Container 1: KCl | Container 2: sand+KCl | Container 3: sand |
|---|---|---|
| $(1.15 \pm 0.004 \pm 0.09) \times 10^{-2}$ | $(3.08 \pm 0.01 \pm 0.26) \times 10^{-3}$ | $(5.37 \pm 0.02 \pm 0.43) \times 10^{-3}$ |

The background subtraction due to the radiation emitted by the laboratory floor, is not straight forward since the detector is shielded by the sample material. Therefore, a reduction factor due to shielding should be applied to the count rate value obtained by placing the detector directly on the laboratory floor. The correct determination of this factor is not trivial, since the radiation reaching the detector comes from several directions. The easiest and most accurate computational method is Monte Carlo simulation, in which two different situations are considered. In the first one, the entire setup is considered. In a second situation, the simulation is repeated by replacing the sample material (KCl, sand+KCl or sand) with air. In both situations the radiation source is a flat circular source with appropriate radius (we took half of the laboratory width). This source emits 1460 keV photons isotropically. Again, only photons with fight lines crossing the detector crystal are followed. The ratio $F_{bgd}$ between the number of photons reaching the detector in the first situation to the number of photons that reach the detector in the second situation provides the attenuation factor. Table 4 presents the values of the attenuation factors obtained for each of the situations. Quoted uncertainties are only statistical.

Table 4: Attenuation factor of 1460 keV radiation to be applied to measure the background count rate, for each experimental situation.

| Container 1: KCl | Container 2: sand+KCl | Container 3: sand |
|---|---|---|
| (0.835 ± 0.014) | (0.397 ± 0.007) | (0.465 ± 0.008) |

## 4. Data analysis

The $^{40}$K nuclide has half-life $T_{1/2} = (1.2504 \pm 0.0030) \times 10^9$ year and the 1460 keV photon emission a branching ratio of $B_R = 0.1055 \pm 0.0011$ [8]. Figure 4 shows a KCl source energy spectrum, where the 1460 keV peak is marked. This peak is easily identified in the spectrum, so no energy calibration is necessary. The spectra obtained for other samples are similar, but peaks have lower amplitude.

Once identified the 1460 keV peak position, the integral was calculated taking into account background subtraction. This is the background mainly due to radiation from other decays and cosmic radiation crossing the detector. The background count rate is fitted with a function of the type $f_{bgd}(x) = a \times x^{-b}$ where $x$ is the MCA channel, and $a$ and $b$ are the fit constants. In figure 5 the result of that fit is presented. The integral value of the signal is then computed as

$$S = \sum_{x_1}^{x_2} C(x) - f_{bgd}(x) \qquad (3)$$

where *C(x)* is the count rate. The summation is carried out between channels $x_1$ and $x_2$ determined by the following process: starting from the channel with the maximum peak value, we look to the left and right of this channel for the channel that first obeys the condition $C(x) - f_{bgd}(x) \leq 0$. Table 5 presents the values of the integrated count rates for each acquisition.

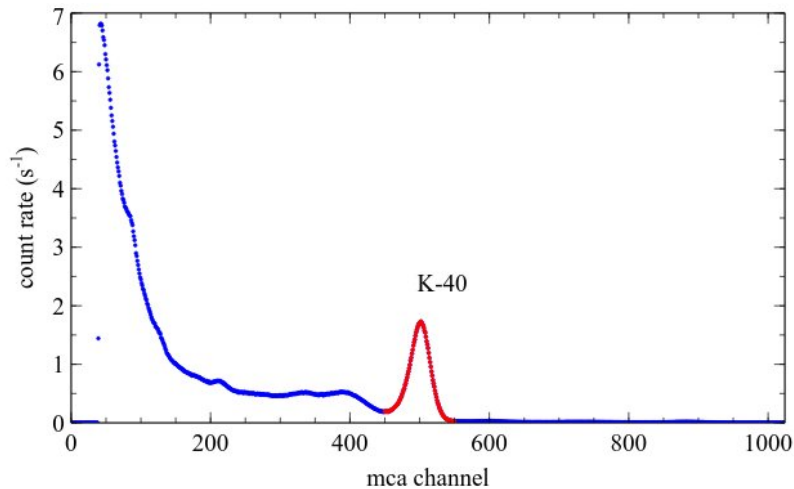

Figure 4. Energy spectrum obtained for container 1 where the 1460 keV peak (red zone) is easily identified. The marked zone corresponds to the peak integration range.

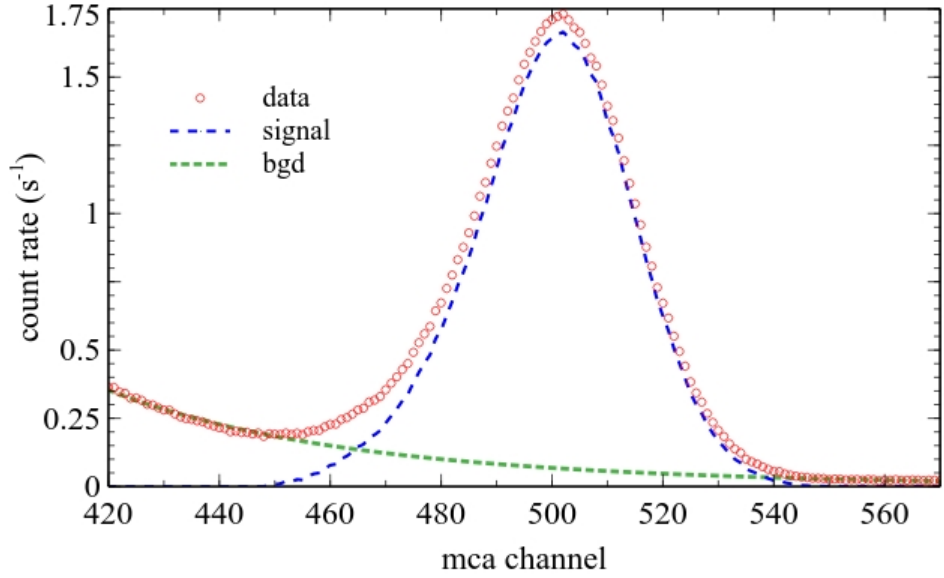

Figure 5. Peak region showing the fit made to obtain the background and the resulting signal curve.

Table 5: Integrated count rates. Uncertainties are only statistical.

| BGD (floor) ($s^{-1}$) | Container 1: KCl ($s^{-1}$) | Container 2: sand+KCl ($s^{-1}$) | Container 3: sand ($s^{-1}$) |
|---|---|---|---|
| 8.25 ± 0.01 | 58.24 ± 0.06 | 48.87 ± 0.05 | 12.74 ± 0.02 |

Using these values, we can calculate activity values for each of the containers using the equation

$$A = \frac{S/t - B/t \times F_{bgd}}{\varepsilon B_r} \qquad (4)$$

where *S/t* is the count rate given in table 5 for each container and *B/t* the count rate due to floor background. Applying equation 4 to each case we obtain the activity values presented in table 6. The uncertainty in the total detection efficiency was obtained by summing in quadrature the statistical and systematic uncertainties presented in table 3. The uncertainty in the activity is then obtained by applying the general uncertainty propagation rules [9].

Knowing the mass of potassium in each of the containers, it's possible to calculate the expected activity due to the isotope $^{40}$K using the equation [10]

$$A = \frac{N_{K40} \ln(2)}{T_{1/2}} \qquad (5)$$

where $N_{K40}$ is the number of atoms of the isotope in the sample and $T_{1/2}$ its half-life. On the other hand, the number of $^{40}$K atoms in each sample can be determined from the mass m of the sample, the average abundance $A_b = (1.17 \pm 0.01) \times 10^{-4}$ of the $^{40}$K isotope in a natural potassium sample and the molar mass $m_A = 74.5513$ g/mol of KCl [11], using the equation [10]

$$N_{K40} = \frac{m \times A_b \times N_A}{m_A} \qquad (6)$$

where $N_A$ is Avogadro's number.

Table 6: Activity according to equation 4 (experimental activity) and calculated according to equation 5 based on known amounts of KCl.

|  | Container 1: KCl (kBq) | Container 2: sand (kBq) | Container 3: sand+KCL (kBq) | Container 3: KCl (kBq) |
|---|---|---|---|---|
| Experimental (eq. 4) | 42.3 ± 3.3 | 15.7 ± 1.2 | 141 ± 12 | 125 ± 12 |
| Computed (eq. 5) | 43.1 ± 0.4 |  |  | 124 ± 1 |

**Conclusion**

The described experiment allows for a non-standard determination of a radioactive source activity. It provides a more complex situation other than the point-like source but where physical quantities can still be computed. The experimental setup can be built with materials with easily accessible. In particular, KCl is affordable and can be bought without significant restrictions, in generous quantities. The needed detection system is rather standard in university laboratories. The choice of Monte Carlo program is not critical and any of the general purpose freely available programs for radiation tracking (like GEANT4 [12,13], FLUKA [14, 15], etc.) can also be used.

Concerning the experimental results, a very good agreement between computed and measured activity values has been obtained. It was found that this agreement is only possible after a careful subtraction of the floor background and taking into account the attenuation introduced by the materials in the containers. The experiment is highly pedagogical, allowing students to put to test their computational skills in the determination of a physical quantity.

**Acknowledgements**

We thank Prof. Alcides Pereira from University of Coimbra for carrying out the elemental analysis of the sand. We also thank Ashley Rose Peralta for proofreading the English text.